\documentclass[aps,prx,twocolumn,preprintnumbers,amsmath,amssymb,superscriptaddress]{revtex4-2}
\usepackage{epsfig}
\usepackage{subfigure}
\usepackage{graphicx}
\usepackage{dcolumn}
\usepackage{bm}
\usepackage{color}
\usepackage[normalem]{ulem}
\usepackage[colorlinks=true,citecolor=blue,linkcolor=blue,urlcolor=blue,hyperindex,CJKbookmarks]{hyperref}\usepackage{amsmath,amsfonts,amsthm,bm, amssymb}
\usepackage{hyperref}
\usepackage{comment}
\usepackage{physics}
\begin{document}
\title{Detecting Thermodynamic Phase Transition via Explainable Machine Learning of Photoemission Spectroscopy} 	
\date{\today}

\begin{abstract}
\textbf{Abstract:} 
Identifying thermodynamic signatures of electronic phases, such as superconductivity, is challenging in low-dimensional materials due to strong fluctuations and low probing volume. Spectroscopic methods are often used to identify new bulk phases, but their main measurable quantity -- electronic energy gaps -- is no longer an effective order parameter in low dimensional and fluctuating systems. Combining angle-resolved photoemission with a domain-adversarial neural network, we report a data-driven method to identify thermodynamic phase transitions solely based on single-particle spectra. We demonstrate 97.6\% accuracy in cuprate superconductor Bi$_2$Sr$_2$CaCu$_2$O$_{8+\delta}$ with strong superconducting fluctuations. This model notably compensates for the scarcity of experimental data by leveraging virtually inexhaustible simulated data. Further, its explainability reveals the crucial role of in-gap spectral weight in detecting phase fluctuations and thermodynamic transitions. Our work pinpoints the spectroscopic signatures of fluctuating orders and enables using spectroscopy for machine-learning-assisted material discovery for low-dimensional and strong coupling systems.
\end{abstract}

\author{Xu Chen}\thanks{X.C. and Y.S. contributed equally to this work.}
\affiliation{Department of Chemistry, Emory University, Atlanta, GA 30322, United States}
\author{Yuanjie Sun}\thanks{X.C. and Y.S. contributed equally to this work.}
\affiliation{Department of Physics and Astronomy, Clemson University, Clemson, SC 29631, United States}
\author{Eugen Hruska}
\affiliation{Department of Chemistry, Emory University, Atlanta, GA 30322, United States}
\author{Vivek Dixit}
\affiliation{Department of Physics and Astronomy, Clemson University, Clemson, SC 29631, United States}
\author{Jinming Yang}
\affiliation{Department of Physics, Yale University, New Haven CT, 06511, United States}
\author{Yu He}
\email{yu.he@yale.edu}
\affiliation{Department of Applied Physics, Yale University, New Haven CT, 06511, United States}
  
\author{Yao Wang}
\email{yao.wang@emory.edu}
\affiliation{Department of Chemistry, Emory University, Atlanta, GA 30322, United States}
\affiliation{Department of Physics and Astronomy, Clemson University, Clemson, SC 29631, United States}

\author{Fang Liu}
\email{fang.liu@emory.edu}
\affiliation{Department of Chemistry, Emory University, Atlanta, GA 30322, United States}

\maketitle

\section{Introduction}
\begin{figure}[!t]
\begin{center}
\includegraphics[width=\columnwidth]{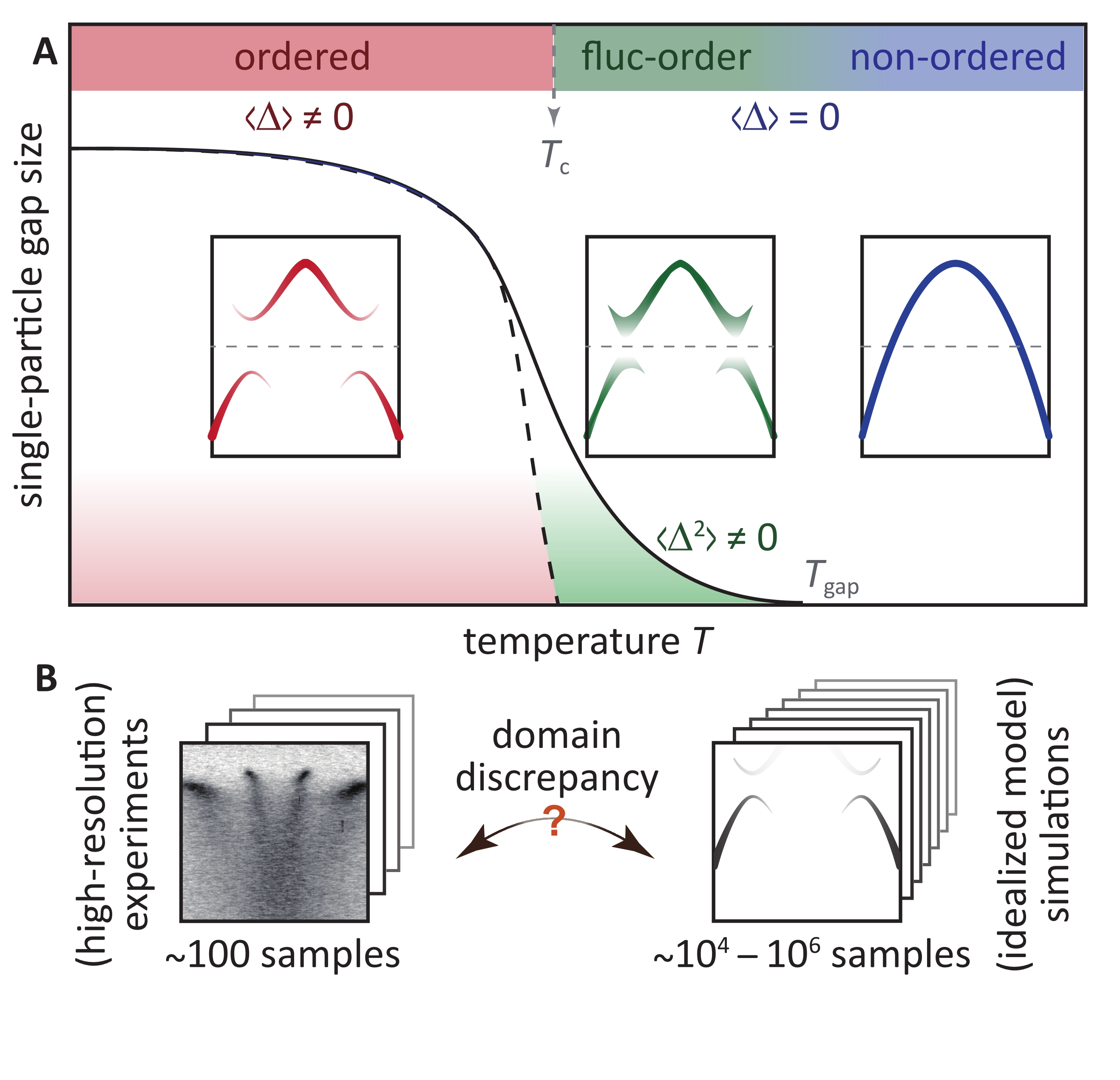}\vspace{-3mm}
\caption{\label{fig:flucCartoon}\textbf{Schematic illustration of thermodynamic phase transition with pronounced fluctuations.} \textbf{A} At low temperatures (red), the material exhibits a nonzero order parameter $\langle \Delta\rangle \neq 0$ and a well-defined single-particle gap. As the temperature increases above $T_c$ (green), the local excitations lose long-range coherence and the average order parameter $\langle \Delta\rangle = 0$. However, the system still displays pronounced fluctuations with a non-negligible $\langle \Delta^2\rangle$, leading to a finite single-particle gap. These short-range fluctuations gradually diminish with further temperature increase, ultimately resulting in a normal state (blue). \textbf{B} The domain discrepancy between the scarce, noisy experimental data and the abundant, idealized simulated spectra. \vspace{-3mm}
}
\end{center}
\end{figure}
New states of matter are typically identified by the emergence of long-range order via a thermodynamic phase transition. As the system crosses the critical temperature $T_c$, singularities in properties like correlation length and specific heat clearly delineate the transition between the ordered (broken-symmetry) phase and the normal (high-symmetry) phase. Traditional thermodynamic probes, such as calorimetry and dilatometry, rely on bulk material measurements and generally lack spatial resolution, making them unsuitable for the characterization of thin van der Waals materials or wafer-scale epitaxial materials. In contrast, electron or optical spectroscopy can be employed to measure the energy gap $\Delta$ (Fig.~\ref{fig:flucCartoon}\textbf{A}), which serves as an indicator of long-range order under the mean-field approximation\,\cite{damascelli2003angle}. Unlike thermodynamic probes, spectroscopy offers the advantages of being non-contact, compatible with \textit{in situ} and operando measurements, and is able to resolve down to micron-level spatial precision in a high-throughput manner\,\cite{boschini2024time, iwasawa2023quantitative}. 
However, many low-dimensional and correlated materials exhibit significant quantum fluctuations in the normal state\,\cite{kondo2015point,he2021superconducting,chen2022unconventional,chen2023role}. Here, the global symmetry is preserved ($\langle \Delta \rangle=0$), but the spectral gap already opens ($\langle \Delta^2 \rangle \neq 0$) which severely undermines the effectiveness of spectroscopy in distinguishing phases\,\cite{sobota2021angle,keimer2015quantum}. In these situations, traditional spectroscopic approach obtains the gap-opening temperature (denoted as $T_{\rm gap}$), which may deviate substantially from the actual thermodynamic transition\,\cite{wang2012interface, ge2015superconductivity, kondo2015point,faeth2020incoherent, xu2020spectroscopic,he2021superconducting,chen2023role}.

Superconductivity (SC) in low-dimensional materials serves as a prime example of how fluctuations can complicate the identification of thermodynamic transitions through spectroscopy. In a well-defined SC phase, global phase coherence among Cooper pairs eliminates electrical resistance, establishing long-range order. Under mean-field approximations, the emergence of the SC phase is linked to a single-particle gap that is generally twice the size of the order parameter. Thus, in scenarios where direct transport measurements are not feasible -- such as in ultrathin films, functionalized surfaces, nonequilibrium systems, or extreme conditions -- the presence of this gap often serves as a fingerprint of SC.\,\cite{lee2014interfacial,wang2012interface, yilmaz2022spectroscopic}. However, in correlated materials like cuprates and monolayer FeSe, quantum and thermal fluctuations disrupt the straightforward relationship between the energy gap and the SC phase, causing a pronounced separation between $T_{\rm gap}$ and $T_c$\,\cite{kondo2015point,faeth2020incoherent, xu2020spectroscopic,he2021superconducting}. Intriguingly, recent studies suggest that angle-resolved photoemission spectroscopy (ARPES) could provide insights into SC phase coherence\,\cite{he2021superconducting,chen2022unconventional, cho2019strongly}, implying that electronic spectra encode many-body information far beyond mean-field descriptions. However, extracting this information remains extremely challenging, requiring material-specific experimental setups, extensive measurements, and detailed temperature-dependent analysis of quasiparticle spectra. The difficulty reflects the pressing need to develop a general method capable of reliably probing thermodynamic phase transitions and the spectral signatures of fluctuating orders with limited experimental data.

Here, we develop a machine learning (ML) model to directly classify actual thermodynamic phase transitions from individual ARPES spectra without needing extensive experimental data or temperature-dependent analysis. One of the core obstacles in applying artificial intelligence (AI) to materials science is the scarcity of labeled experimental data, a ubiquitous issue across scientific fields\,\cite{fagnan2019data}. While most modern ML techniques rely heavily on large datasets, ARPES spectra are particularly challenging to acquire in sufficient quantities due to the inherent complexity of the experiments and the lack of standardized data curation processes. One potential solution is to generate training data through simulations, enabling efficient exploration of parameter space \textit{in silico}\,\cite{ratner2019bes}. However, simulated data, even when augmented with synthetic noise, fall short of replicating real-world experimental noise, resolution constraints, and lab-specific variability (see Fig.~\ref{fig:flucCartoon}\textbf{B}), leading to poor performance when applying simulation-trained models to actual experimental data. To address this experiment-simulation discrepancy, we employ an adversarial training strategy to ensure that the ML model learns classification rules transferable between simulated and experimental ARPES spectra. By leveraging this approach, our model accurately classifies SC states without needing labeled experimental data during training. Furthermore, occlusion-based attribution analysis reveals that the spectral weight distribution near the Fermi level, rather than simply detecting the presence of a gap, is key in distinguishing between SC and normal phase. This approach provides a pathway for integrating simulated single-particle functions with experimental ARPES spectra via ML, with broader applications in the study of thermodynamic phase transitions in quantum materials where labeled experimental data are scarce.

\section{Results}

\begin{figure*}[!t]
\begin{center}
\includegraphics[width=18cm]{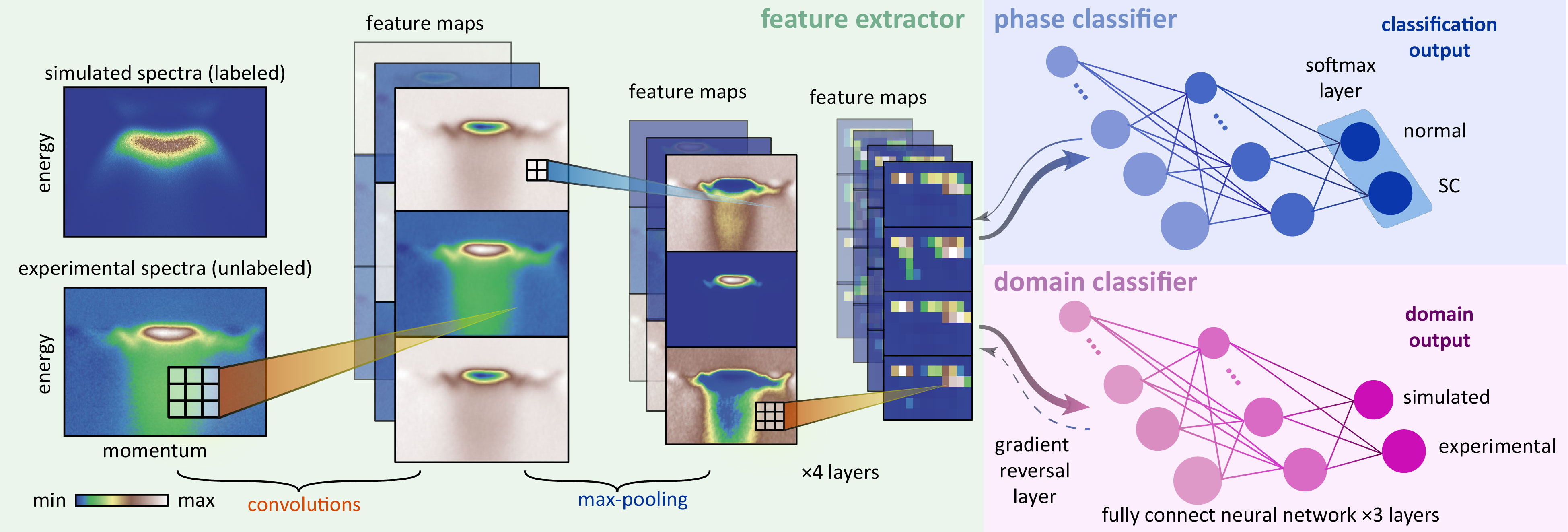}\vspace{-3mm}
\caption{\label{fig:overview} 
\textbf{Model architecture.} Left to right: Consecutive convolutional stacks as the feature extractor convert the ARPES spectrum (using experimental data for BSCCO OD58 as an example) into feature maps with various convolutional kernels. A pooling layer and activation are then applied to compress the feature maps and pass the data to the next layer. After four convolutional layers, the feature maps are pooled and flattened and then directed into two distinct fully connected neural networks: the phase classifier (blue) classifies the spectra into the superconducting (SC) and normal phases; The domain classifier (pink) classifies the sample source either from simulation or experiment. It connects to the feature extractor via a gradient reversal layer that multiplies the gradient by a certain negative constant during the backpropagation (dashed line). The number and size of each layer plotted in this figure are for illustrative purposes. The color scale bar (bottom left) indicates the spectral intensity of the input ARPES spectra, ranging from minimum (blue) to maximum (white).
}
\end{center}
\end{figure*}
\subsection{Model setup and ARPES data curation}
A typical ARPES spectrum collected along a fixed momentum cut is shown in Fig.~\ref{fig:flucCartoon}\textbf{B}. The spectral weight distribution along the horizontal and vertical axes reflects the momentum and energy of a single electron inside the material. Convolutional neural networks (CNNs) are specifically suited to analyze the ARPES spectra because the convolution operation can preserve the two-dimensional (2D) structure of momentum-energy information, instead of flattening the spectrum into a vector (see Note S1 and Figure S1 for details).
However, standard CNN models' performance can be significantly hampered by the ``domain shift" issue in our case -- the training data (simulated spectra) and test data (experimental spectra) are from different sources.
Hence, we need to use a modified CNN version, the domain-adversarial neural network (DANN)\,\cite{JMLR:v17:15-239}, to facilitate cross-domain learning. In this work, as illustrated in Fig.~\ref{fig:overview}, we adapt a DANN architecture that consists of three main components: a feature extractor, a domain classifier, and a phase classifier.

The feature extractor, implemented through convolutional layers, extracts feature representations from both simulated and experimental ARPES spectra. These representations are then passed to the phase classifier, which predicts the material phase as either SC or normal. During training, the model parameters of the feature extractor and the phase classifier are optimized by minimizing the phase label classification loss. The additional domain classifier addresses the domain shift issue through an adversarial training process, encouraging the network to discover domain-invariant latent representations shared between simulated and experimental spectra. Specifically, the domain classifier determines whether the extracted features originate from the simulated or experimental data via minimizing the cross-entropy loss, while the feature extractor is trained to confuse the domain classifier (see \hyperref[sec:domainAdp]{domain adaptation} in the methods). 
Upon completion of the DANN training, the model is able to classify any single simulated or experimental ARPES snapshot. 

Notably, to work with limited experimental data, the supervised learning of the phase classifier is entirely guided by simulated spectra with phase labels. However, unlabeled experimental spectra are used to train the domain classifier to ensure that a robust latent representation is learned by the feature extractor. Since the adversarial training process ensures domain transferability, the predicted phase labels for experimental data are expected to reflect the true material phases, even without direct supervision during training.

\begin{figure}[!t]
\begin{center}
\includegraphics[width=\columnwidth]{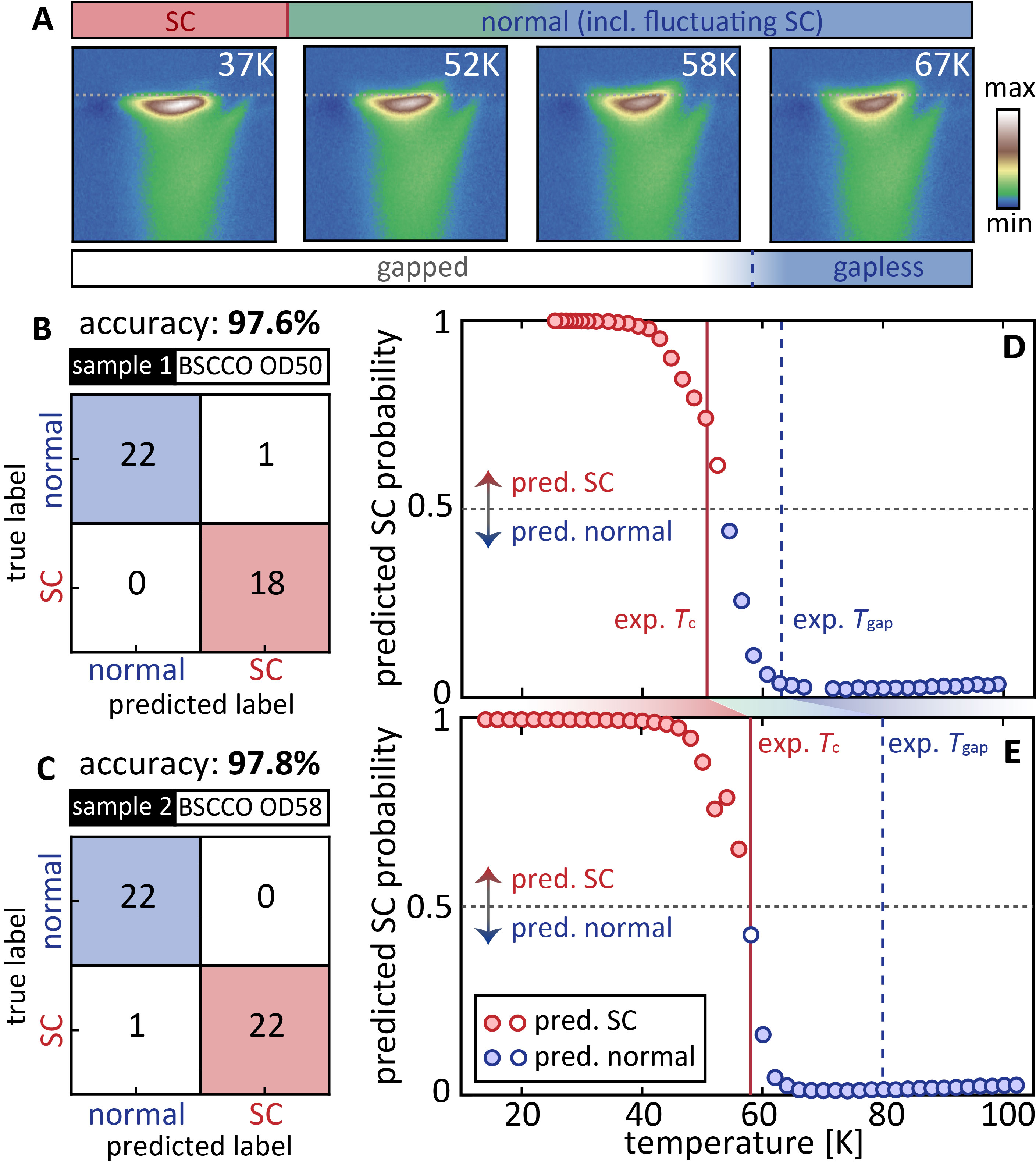}\vspace{-3mm}
\caption{\label{fig:2compRes} \textbf{Classification of superconducting phase using the DANN model.} \textbf{A} Experimental ARPES spectra for the BSCCO OD50 sample at four different temperatures, with dashed lines denoting the Fermi level. The upper and lower bars indicate the phase labels, independently determined by experiments and invisible to the ML model, and the gap sizes. The color scale bar indicates spectral intensity, ranging from minimum (blue) to maximum (white).\textbf{B},\textbf{C} Confusion matrices obtained for binary classification of the ARPES spectra collected from \textbf{B} BSCCO OD50 and \textbf{C} OD58 samples at various temperatures, yielding accuracies of 97.6\% and 97.8\%, respectively. \textbf{D},\textbf{E} The ML-predicted SC probability $\bar{p}_{\textrm{SC}}$ for spectra obtained from the \textbf{D} BSCCO OD50 and \textbf{E} OD58 samples, respectively. The $\bar{p}_{\textrm{SC}}$s are calculated for each spectrum but are sorted here by their experimental temperatures, which are unknown to the ML model. The classification is based on whether $\bar{p}_{\textrm{SC}}$ exceeds 0.5. Correctly classified spectra are denoted by solid red and blue dots, while misclassified data are depicted by open dots. The red and blue lines indicate experimentally determined transition temperature $T_c$ and gap-opening temperature $T_{\rm gap}$ (both invisible to the ML model), respectively.\vspace{-2mm}
}
\end{center}
\end{figure}

The experimental ARPES spectra were collected from two cuprate samples: a super-oxygen-\linebreak ated Bi$_2$Sr$_2$CaCu$_2$O$_8$ (BSCCO) with $T_c = 50$\,K (OD50) and an oxygenated BSCCO with $T_c = 58$\,K (OD58) (see details of experimental determination of $T_c$ in Note S6). A total of 41 and 45 spectra are measured in these two samples, respectively, spanning a temperature range from 14\,K to 102\,K, covering both SC and normal phases (reused from He et al.\,\cite{he2021superconducting, he2018rapid}). Superconducting gap extraction with traditional spectral fitting method is conducted to benchmark the performance and advantage of the new method here (see Note S5 and Figure S7, and S8). All labeled training data are generated through simulations following the method outlined in \hyperref[sec:simulation]{simulated ARPES data}. A uniform background removal process is applied to both the simulated and experimental datasets (see supplemental methods and Figure S5 for details). Fig.~\ref{fig:2compRes}\textbf{A} presents four spectra taken from the OD50 sample after the background removal. Notably, while superconductivity disappears when the temperature exceeds $T_c=50$\,K, the single-particle gap remains open until $T_{\rm gap}\sim 65$\,K\,\cite{he2018rapid}.

\subsection{Superconducting phase classification}
The trained DANN model is applied to classify superconducting phases in BSCCO OD50 and OD58 experimental spectra, where the pronounced fluctuating gap long hinders the spectral characterization of superconducting phase with traditional spectroscopy methods\,\cite{he2021superconducting}. As shown in Figs.~\ref{fig:2compRes}\textbf{B} and \ref{fig:2compRes}\textbf{C}, the model's predictions are represented by a scalar ``SC probability'' $\bar{p}_{\textrm{SC}}$ for each input spectrum. Such an SC probability $\bar{p}_{\textrm{SC}}$ is obtained from an ensemble approach, where $\bar{p}_{\textrm{SC}}$ for each spectrum is the averaged $p_{\textrm{SC}}$ over 10 different DANN models with various initializations. Normalized by the sum rule, a sample corresponding to an input spectrum is predicted as SC when $\bar{p}_{\textrm{SC}}$ exceeds 50\%. For the BSCCO OD50 sample, only one spectrum measured in the normal phase is misclassified as SC. Conversely, all other 40 experimental spectra are correctly classified, yielding an accuracy of 97.6\%. The performance is similar for the OD58 sample, where only one spectrum in the SC state is misclassified. 

As the experimental spectra are obtained for two material samples at different temperatures, we further analyze the predicted SC probability $\bar{p}_{\textrm{SC}}$ as a function of temperature in Figs.~\ref{fig:2compRes}\textbf{D} and \ref{fig:2compRes}\textbf{E}. Although our ML model independently classifies each ARPES spectrum and does not have access to temperature information, we observe a roughly monotonic relationship between $\bar{p}_{\textrm{SC}}$ and the actual temperature $T$. The overall shape of the $\bar{p}_{\textrm{SC}}(T)$ resembles an inverted sigmoid function: it shows constantly high values (close to 1) at low temperatures and low values (close to 0) at high temperatures, with a rapid transition from 1 to 0 near the experimentally determined $T_c$. The only misclassified data point in each of Figs.~\ref{fig:2compRes}\textbf{D} and \ref{fig:2compRes}\textbf{E} corresponds to a sample measured at a temperature close to $T_c$. Given that none of the experimental spectra are labeled in ML, the monotonic decrease in the predicted probability with increasing temperature indicates that our model has successfully captured the thermodynamic phase transition encoded in the details of ARPES spectra. Interestingly, the predicted $\bar{p}_{\textrm{SC}}$ does not exhibit any anomaly near the $T_{\rm gap}$. This is in stark contrast to traditional superconducting spectral gap fitting results, which is instead only sensitive to $T_{\rm gap}$, and furthermore suffers a rapidly exacerbating stability issue above $T_c$ (see Note S5 and Figure S7-S8). These observations reflect that the gap opening is not the sole or even primary indicator of a thermodynamic phase transition in many quantum materials. Instead,  spectral distributions beyond the band dispersion contain extensive information about the emergence of long-range order associated with this phase transition.
\begin{figure}[!t]
\begin{center}
\includegraphics[width=\columnwidth]{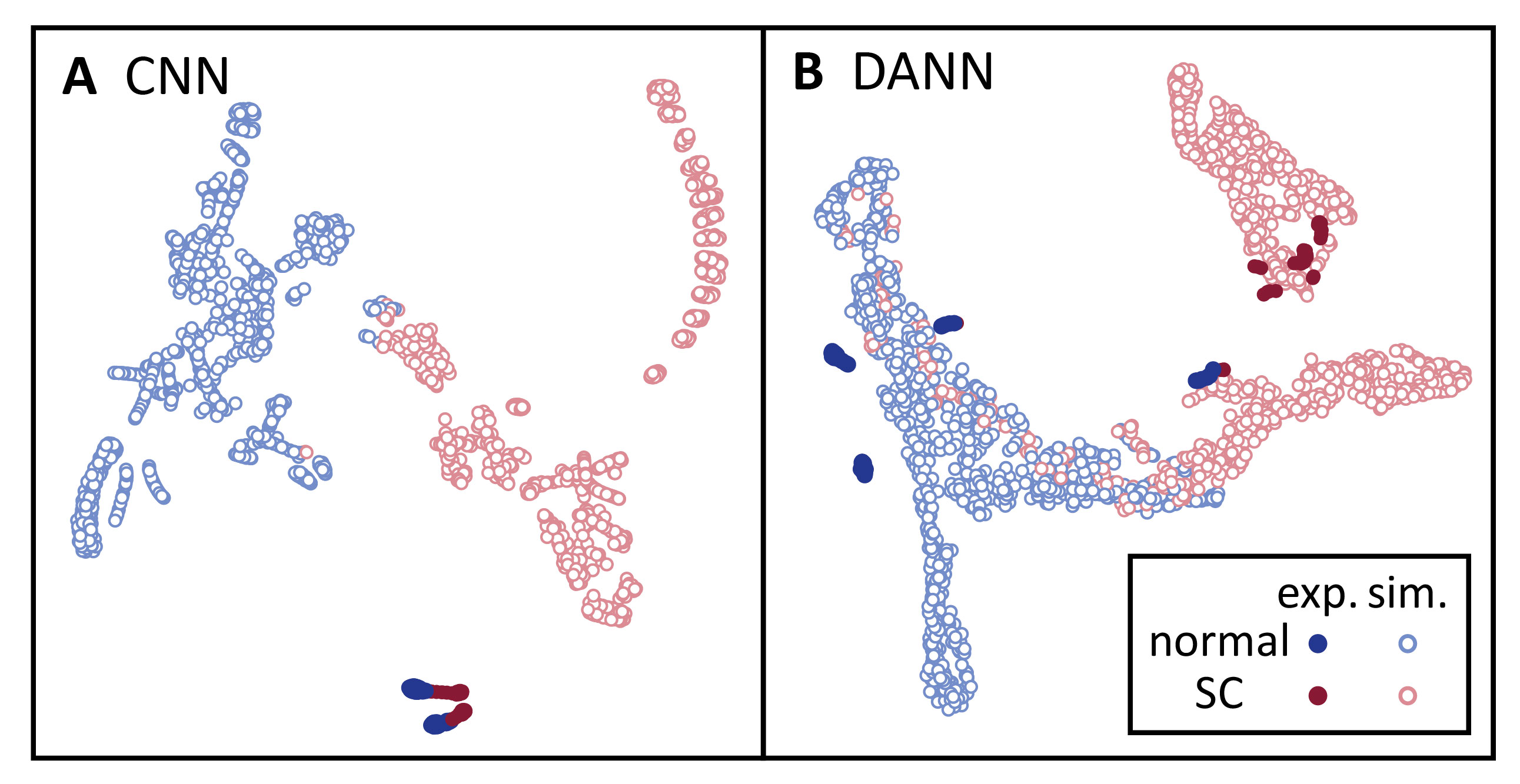}\vspace{-3mm}
\caption{\label{fig:tsne} \textbf{t-SNE visualization of feature distributions.} \textbf{A} The distribution of the feature extractor's activations of the CNN model without domain adaptation. \textbf{B} The distribution of feature extractor's activations of the CNN model when the adaptation procedure was incorporated into training. Light blue and red colors indicate the simulated spectra belonging to normal or SC phases, respectively, while dark colors represent experimental spectral data.
}
\end{center}
\end{figure}

It is worth noting that the simulated and experimental ARPES spectra exhibit fundamental differences, as shown in Fig.~\ref{fig:flucCartoon}\textbf{B}, due to the simplicity of the single-band model and unreplicable experimental noise. Thus, the domain adaptation technique is crucial in bridging the gap between the simulated training spectra and experimental test spectra, hence ensuring our ML phase classifier's high accuracy.  Without the adversarial training enabled by the domain classifier, the CNN model yields a representation space where the simulated and experimental spectra occupy distinct regions of the latent space, as visualized by the t-distributed stochastic neighbor embedding\,\cite{JMLR:v9:vandermaaten08a} (t-SNE) in Fig.~\ref{fig:tsne}\textbf{A}. This separation reflects the intrinsic differences between simulated and experimental data in this representation. As a result, the phase classification rules learned from the simulated spectra are not directly applicable to experimental spectra. In contrast, the domain classifier in DANN ensures a better representation is learned to capture the underlying physics rather than the apparent differences, significantly enhancing the alignment between simulated and experimental spectra in the representation space (see Fig.~\ref{fig:tsne}\textbf{B}).  Consequently, the DANN provides transferable phase classification rules between simulated and experimental spectra, and the classification results show a substantial improvement by 19.0\% and 19.2\% in average accuracy on the BSCCO OD50 and OD58 samples, respectively (see Note S2 and Tables S1 and S2). 

\begin{figure}[!t]
\begin{center}
\includegraphics[width=\columnwidth]{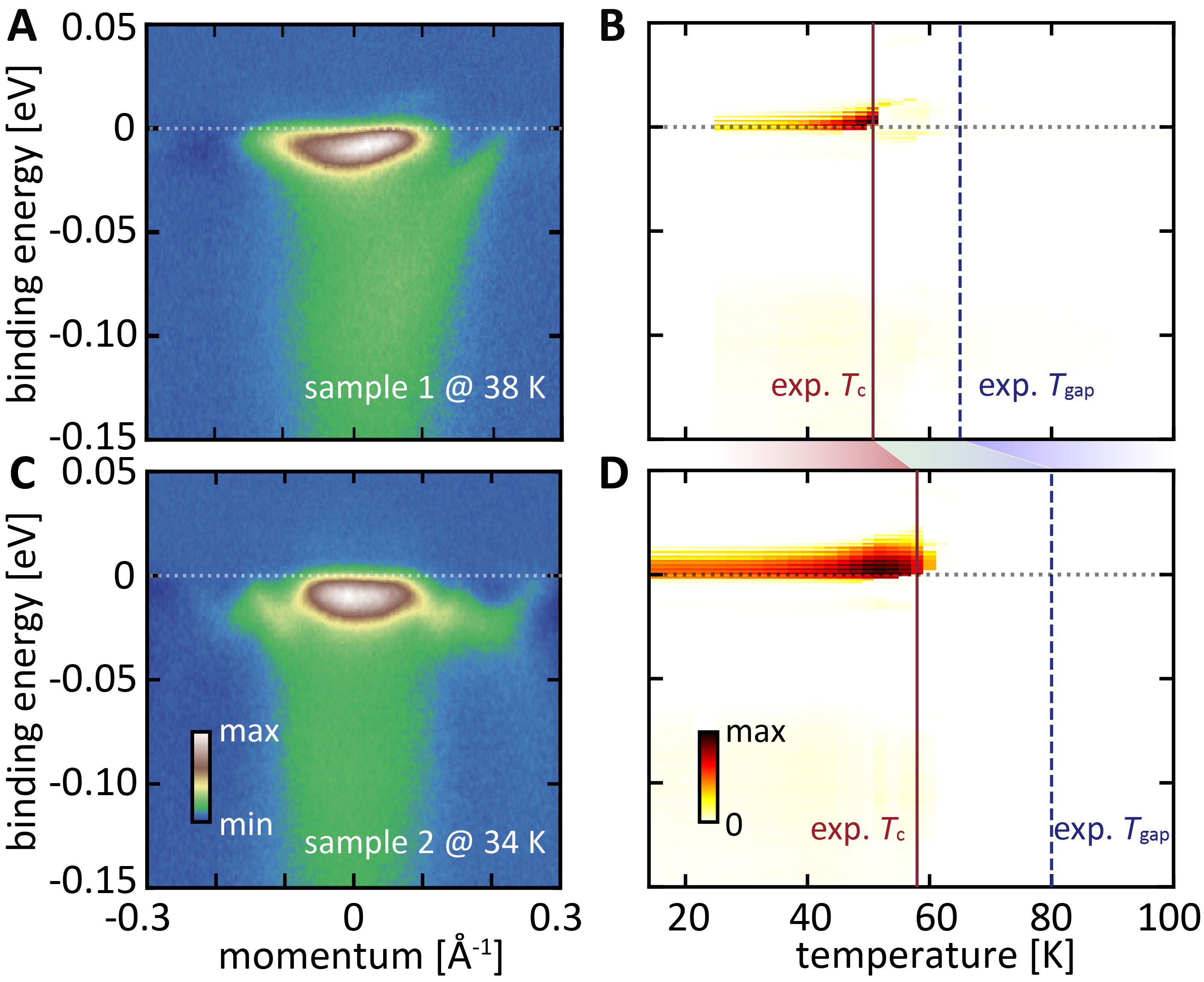}\vspace{-3mm}
\caption{\label{fig:saliency} \textbf{Saliency distribution for experimental spectra.} \textbf{A} A sample ARPES spectrum obtained from BSCCO OD50 at 38 K (superconducting phase). The color scale bar indicates the spectral intensity, ranging from minimum (blue) to maximum (white). \textbf{B} The saliency values across various binding energies and temperatures for the BSCCO OD50 sample, quantifying the sensitivity of the ML prediction for each energy. The color scale bar indicates positive saliency, from low (white) to high (black). \textbf{C} An ARPES spectrum from BSCCO OD58 at 34 K (superconducting phase). The color scale bar represents the spectral intensity, as in (A). \textbf{D} Saliency values for the BSCCO OD58 sample. The color scale is the same as in (B). Grey dashed lines denote the Fermi level, while red and blue lines indicate the $T_c$ and $T_{\rm gap}$, as independently determined by experiments.
}
\end{center}
\end{figure}

\subsection{Physical explanation of the ML model}
Beyond classifying SC from ARPES spectra, we also aim to extract interpretable physical intuitions from the ML model to deepen our understanding of quantum materials and, especially, to identify single-particle spectral features directly linked to a thermodynamic phase transition. Specifically, we exploit an occlusion-based analysis for the ML model to identify which single-particle spectral features are linked to the SC long-range order\,\cite{DBLP:journals/corr/ZeilerF13}. The occlusion-based analysis measures the impact of a specific data feature on a model's output when that feature gets blocked. In the context of analyzing ARPES spectra, the occlusion is realized by blocking a region of the spectral function information via an energy-resolved occluding patch.

The occluded spectrum, $\tilde{A}(\mathbf{k},\omega;\nu)$, of the original spectrum, $A(\mathbf{k},\omega)$, is defined as
\begin{equation}\label{eq:occludedAkw}
\tilde{A}(\mathbf{k},\omega;\nu) = A(\mathbf{k},\omega) + \delta(\omega-\nu)[A_0(\mathbf{k},\nu) - A(\mathbf{k},\nu)]\,.
\end{equation}
\noindent
Here, $A_0(\mathbf{k},\nu)$ is a single baseline spectrum at energy $\nu$, which provides the spectral function value in the occluded region. Following recent feature attribution studies\,\cite{sturmfels2020visualizing, erion2021improving}, we set $A_0(\mathbf{k},\nu)$ as the training distribution to ensure robust explanation performance across different pixel intensities (see details in supplemental methods and Figure S4).

The change in the predicted probability before and after occlusion is usually referred to as saliency. By shifting the occluding patch in Eq.~\eqref{eq:occludedAkw} along the energy axis, we can determine the saliency for each spectrum sample as a function of energy $\nu$:
\begin{equation}\label{eq:saliency}
S_{\text{SC}}(\nu) = \max \left\{
0, \bar{p}_{\text{SC}}[A(\mathbf{k}, \omega)] - \bar{p}_{\text{SC}}[\tilde{A}(\mathbf{k}, \omega; \nu)] 
\right\}\,,
\end{equation}
\noindent
where $\bar{p}_{\textrm{SC}}[\cdot]$ represents the ML model mapping from an input ARPES spectrum to the predicted SC probability $\bar{p}_{\textrm{SC}}$. To focus on features that positively contribute to the classification, negative saliency values are filtered out by a rectified linear unit (ReLU) function\,\cite{agarap2018deep} (the max operation in Eq.\ref{eq:saliency}). 

As shown in Fig.~\ref{fig:saliency}, we obtain the saliency as a function of binding energy for each spectrum in both BSCCO OD50 and OD58 experimental datasets. A notable feature across various spectral samples is a pronounced peak at zero energy ($\nu=0$). Remarkably, the saliency magnitude significantly decreases for spectra acquired at temperatures above $T_c$. This observation suggests that the ML model identifies the nuanced distribution of spectral weight near and above the gap center as the key for pinpointing the true $T_c$. The spectral weight just above the gap center is particularly sensitive to the upper branch formation of the Bogoliubov quasiparticles, whose clear separation from the lower branch is a major empirical identifier of superconducting phase ordering in cuprates\,\cite{he2021superconducting}. In an intriguing parallel, a recent study inferred electronic entropy from a continuous sequence of ARPES spectra, achieved through meticulous temperature and energy calibration\,\cite{chen2022unconventional}. This study deduced that the temperature-derivative of weighted in-gap spectral intensity near the Fermi level could effectively determine the $T_c$. Our ML model's explanation aligns with this conclusion, and also naturally points to the minimal experimental requirement for our method to work: enough resolution and signal-to-noise to resolve in-gap spectral weight. However, different from the temperature-derivative approach, the ML classification and the saliency analysis operate on each individual spectrum, without relying on temperature dependent information. The experimental data, used solely for domain alignment, remain unlabeled. This approach of single-snapshot-based classification approach is particularly valuable for general material design, where exhaustive parameter tuning and sequential measurements are impractical, such as under extreme or non-equilibrium conditions.

\subsection{Ternary state classification}
We now turn to investigate the fluctuating superconducting (fluc-SC) regime within the normal state. This regime is characterized by pronounced short-range Cooper pair fluctuations, manifesting as a single-particle gap comparable to that observed below $T_c$\,\cite{kondo2015point,faeth2020incoherent, xu2020spectroscopic,he2021superconducting}. To further assess the robustness of our model and method, we expand the classification task to include three states: SC, fluc-SC, and gapless non-SC states. Specifically, we re-label simulated spectral data to reflect these three states and retrain the DANN model, while keeping the experimental data unlabeled. This model is then applied to individual ARPES experimental spectra to distinguish whether the corresponding sample material is in the SC, fluc-SC, or gapless non-SC state.

\begin{figure}[!t]
\begin{center}
\includegraphics[width=0.8\columnwidth]{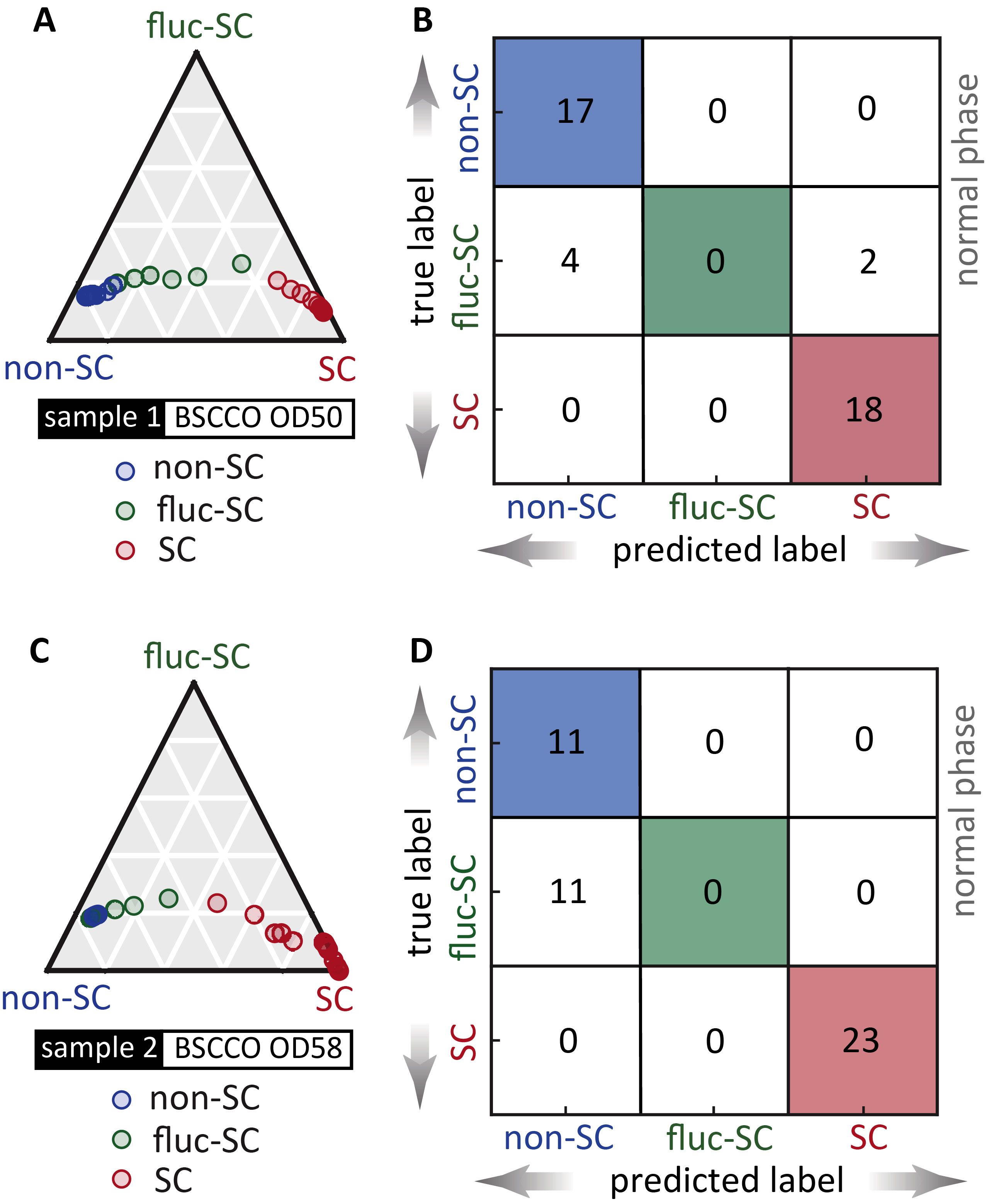}\vspace{-4mm}
\caption{\label{fig:3comp} \textbf{Ternary classification results for two BSCCO samples.} \textbf{A} Predicted probability in the ternary graph for superconducting (red), fluctuating-superconducting (green), and gapless non-superconducting (blue) phases using the DANN model for experimental spectral data measured in the BSCCO OD50 sample. The proximity of each point to each vertex represents the probability for the corresponding phase, while the color reflects its true label determined by experiments. \textbf{B} Confusion matrix for the ternary classification in \textbf{A}. \textbf{C},\textbf{D} Same as \textbf{A} and \textbf{B} but for BSCCO OD58 samples. 
}
\end{center}
\end{figure}
As shown in Fig.~\ref{fig:3comp}, the spectra stemming from SC and gapless non-SC states are still correctly classified under this ternary classification. However, distinguishing the fluc-SC phase presents greater challenges, resulting in an overall accuracy of 85.4\% for the BSCCO OD50 sample and 75.6\% for the OD58 sample. An interesting observation is that the majority (15 out of 17) of gapped fluc-SC states are misclassified as gapless non-SC, despite the presence of a gap, which is traditionally considered a fingerprint of long-range order. Instead, our ternary classification results correctly identify the \textit{inherent similarities} between gapped fluc-SC and gapless non-SC states, both of which exist above $T_c$ in the normal state. This result suggests that traditional single-particle spectroscopy, which relies heavily on gap detection, oversimplifies the complexities fluctuating states to a single gap value. The gapped fluc-SC state and the gapless non-SC state are inherently a crossover, which also proves more difficult to distinguish compared to the detection of true SC long-range (phase) ordering. Although the accuracy decreases for the fluc-SC phase, domain adaptation remains crucial in improving performance. It enhances the accuracy of the phase classification by more than 10\% and 30\% compared to a traditional CNN, with (Table S1) and without (Table S2) ensemble averaging, respectively.

\section{Discussion}
Our study demonstrates that a transferable ML model, trained without labeled experimental data, effectively classifies the SC of strongly correlated materials based on single-snapshot ARPES spectrum. This breakthrough suggests a powerful new tool for identifying thermodynamic phase transitions and long-range orders from electronic spectroscopy, even when the deterministic features do not conform to traditional gap analysis. The model's transferability between simulation and experiment is demonstrated through its accurate classification of ARPES spectra from two materials with distinct compositions, gap sizes, and bare band dispersions and that are measured at a variety of temperatures. This pre-trained model is also expected to accurately classify other superconductive materials similar to cuprates in a traditional inductive manner (see Note S3 and Tables S3 and S4 for details). For materials significantly different from cuprates, accurate classifications can still be achieved through transductive learning by fine-tuning the model with corresponding unlabeled experimental spectra (Note S3). In both cases, our model bypasses the need for tracking temperature-dependent trends, thereby enabling non-contact, \textit{in situ}, and operando characterization of superconductivity in correlated materials, with potential in pump-probe spectra where phases lose coherence\,\cite{boschini2018collapse,yang2018terahertz, sun2020transient}. Furthermore, the physical intuition gained from the ML model, revealed through occlusion-based attribution analysis, closely matches the recent conclusion from temperature-dependent analysis\,\cite{chen2022unconventional}. Our approach fits into the growing emphasis on transfer learning\,\cite{bozinovski2020reminder, pratt1992discriminability} and domain adaptation techniques\,\cite{ben2010theory,han2021transfer} in scientific applications to overcome data scarcity. Despite the unique challenges compared to existing efforts, including fewer available experimental data, a larger discrepancy between simulation and experiment, and more complicated underlying patterns, our model achieves impressive results in classifying superconductivity in quantum materials, with the potential to classify other thermodynamic phase transitions after retraining with new simulated data generated from relevant phenomenological equations. Our model paves the way for high-throughput material design and synthesis experiments that require immediate characterizations\,\cite{yamada1998doping}.

\section{Methods} \label{methods}

\subsection{Simulated ARPES data}\label{sec:simulation}
ARPES data simulations are based on the phenomenological model described in He et al.\,\cite{he2021superconducting}, which applies to a broad range of superconducting materials\,\cite{norman1998, cho2019strongly, sacepe2011localization,sacepe2008disorder}. The spectral function used in the simulation is given by:
\begin{equation}\label{eq:simSpec}
    \begin{split}
        A(\mathbf{k},\omega) =& \frac{E_\mathbf{k}+ \epsilon_\mathbf{k}}{\sqrt{8\pi}\sigma E_\mathbf{k}}e^{-\frac{(\omega+E_\mathbf{k})^2}{2\sigma^2}} + \frac{E_\mathbf{k} - \epsilon_\mathbf{k}}{\sqrt{8\pi}\sigma E_\mathbf{k}}e^{-\frac{(\omega-E_\mathbf{k})^2}{2\sigma^2}}\,,
    \end{split}
\end{equation}
where $E_\mathbf{k} = \sqrt{\epsilon_\mathbf{k}^2+\Delta_\mathbf{k}^2}$ is the Bogoliubov quasiparticle dispersion, $\Delta_\mathbf{k}$ is the momentum dependent superconductivity order parameter, and $\sigma$ is the spectral broadening.

The superconducting order parameter $\Delta_\mathbf{k}$ is determined by the general band BCS gap equation, with an approximate closed-form solution over the entire temperature range:
\begin{equation}\label{eq:simGap}
    \Delta_\mathbf{k}(T) = \Delta_\mathbf{k}(0) \text{tanh}\left(2.34\sqrt{T_\text{gap}/T-1}\right)\,.
\end{equation}
Here, $T_{\rm gap}$ denotes the superconducting pairing onset temperature, while $T_c$ is the thermodynamic transition temperature. Both parameters are randomly sampled in the simulations to enhance model transferability (see Note S4 and Figure S2)\,\cite{he2018rapid,chen2022unconventional}. States between $T_c$ and $T_\text{gap}$ are labeled as fluc-SC. The coefficient 2.34 in Eq.~\eqref{eq:simGap} is specific to $d$-wave superconductors and may slightly vary for different Fermi-surface geometries\,\cite{devereaux1995electronic}. However, as it simply rescales temperature, which remains hidden in DANN training, its value is fixed in the simulated dataset without compromising transferability.

The spectral width $\sigma(\omega, T)$ is modeled to reflect realistic energy and temperature dependencies, ensuring alignment with experimentally observed values (in eV and Kelvin):
\begin{equation}\label{eq:simWidth}
    \sigma(T) = 0.028 + 5.56\times10^{-7}T^2 + 0.01 \text{tanh}[5(T/T_c-1)] + 5\omega^2\,.
\end{equation}
These dependencies account for electronic self-energy within a Fermi-liquid framework\,\cite{damascelli2003angle} and additional broadening due to phase-mode scattering above $T_c$\,\cite{he2021superconducting,kondo2015point,chen2022unconventional}. 
To enhance transferability across different superconducting materials and experimental conditions, the resulting spectra, after applying the theoretical broadening in Eq.~\eqref{eq:simWidth}, undergo convolution with a resolution function, whose parameters are randomly sampled. The energy-momentum resolution is implemented as a 2D Gaussian convolution on the 2D spectral function, with a typical momentum resolution of 10-pixel size (see Note S4). 

The background is treated as 10 times the average per-pixel intensity across the simulated cut multiplied by the Fermi function, which mimics the momentum-scrambling secondary scattering process in photoemission. Since we are mostly in intermediate to high count rate regime, the noise distribution is approximated as a Gaussian with a standard deviation of $\alpha\sqrt{N_{ij}}$, where $N_{ij}$ is the simulated count at pixel $\{i,j\}$, and $\alpha$ is an input parameter to control the signal to noise ratio. This approximation mimics the Poisson noise observed in real experiments, which is generally in the high-count-rate regime and asymptotically approaches a Gaussian noise. The entire spectrum takes an absolute value to eliminate negative counts due to the application of Gaussian noise at extremely low count regimes, which has negligible impact on the main spectral region of interest. A comparison of the parameter space of simulated and experimental data can be found in Note S3. A total of 1,745 simulated spectra were obtained, with 80\% (1,395) being utilized as training data. 

\subsection{Domain adaptation} \label{sec:domainAdp}
DANN is a domain adaptation technique designed to reduce the domain shift between the source (simulated) and target (experimental) domains. The model architecture comprises three parts as shown in Fig.~\ref{fig:overview}: the feature extractor ($G_f(\cdot;\theta_f)$) with parameters $\theta_f$, the phase classifier ($G_y(\cdot;\theta_y$)) with parameters $\theta_y$, and the domain classifier ($G_d(\cdot;\theta_d$)) with parameters $\theta_d$. Training DANN involves minimizing the phase classification loss using labels from the source domain while simultaneously encouraging the feature extractor to learn domain-invariant features through adversarial training (see detailed illustrations of forward and back propagation flow in Figure S6). This is achieved using a gradient reversal layer (GRL) (denoted as $\mathcal{R}(\cdot)$), placed between the feature extractor and the domain classifier. The GRL acts as an identity transformation during forward propagation but flips the sign of $\partial\mathcal{L}_d/\partial\theta_f$ during backpropagation, effectively making the feature extractor maximize the domain classification loss while the domain classifier minimizes it. We define the $\mathcal{L}_y$ and $\mathcal{L}_d$ as the corresponding cross-entropy loss\,\cite{goodfellow2016deep} for label prediction and domain classification, respectively. The general form of the cross-entropy loss for a classification task is:
\begin{equation}
    \mathcal{L}(p, y) = -\sum_{k=1}^{K} y_k \log p_k\,,
\end{equation}
where $K$ is the number of classes; $p$ is the predicted probability vector for each class; $y$ is the one-hot encoded true label vector. Applying this to our specific losses, the phase classification loss for a single sample with true phase label $y \in \{ 0,1 \}$ and predicted probability $\bar{p}_{\textrm{SC}}$ is defined as:
\begin{equation}
\mathcal{L}_y(\bar{p}_{\textrm{SC}}, y) = -y\log(\bar{p}_{\textrm{SC}}) - (1-y)\log(1-\bar{p}_{\textrm{SC}})\,.
\end{equation}
For domain labels $d \in \{ 0,1 \}$ (with $d=0$ for the source domain and $d=1$ for the target domain), and with $\bar{p}_{\textrm{target}}$ being the predicted probability that a sample is from the target domain, the domain classification loss is defined as:
\begin{equation}
\mathcal{L}_d(\bar{p}_{\textrm{target}}, d) = -d\log(\bar{p}_{\textrm{target}}) - (1-d)\log(1-\bar{p}_{\textrm{target}})\,.
\end{equation}
The training objective for the whole DANN is formulated as follows:
\begin{flalign}
  \begin{aligned}
    E(\theta_f,\theta_y,\theta_d) = \mathbb{E}_{(\mathbf{x}_i^s, y_i^s)  \sim \mathcal{D}_s}\mathcal{L}_y\big(G_y(G_f(\mathbf{x}_i^s;\theta_f);\theta_y),y_i^s\big)\\
    +\lambda\Big(\mathbb{E}_{\mathbf{x}_i^s  \sim \mathcal{D}_s}\mathcal{L}_d\big(G_d(\mathcal{R}(G_f(\mathbf{x}_i^s;\theta_f));\theta_d),d^s\big)\\+
    \mathbb{E}_{\mathbf{x}_j^t  \sim \mathcal{D}_t}\mathcal{L}_d\big(G_d(\mathcal{R}(G_f(\mathbf{x}_j^t;\theta_f));\theta_d),d^t\big)\Big)\,,
    \end{aligned}
\end{flalign}
\noindent
where $\mathbb{E}$ represents the expected value; $\mathbf{x}_i^s$ denotes a simulated spectrum with phase label $y_i^s$ from the source domain $D_s = \{ (\mathbf{x}_i^s, y_i^s) \}_{i=1}^{n_s}$; $\mathbf{x}_j^t$ denotes an experimental spectrum in the target domain $D_t = \{ (\mathbf{x}_j^t) \}_{j=1}^{n_t}$; $d^s$ and $d^t$ are the domain labels. Here, $\lambda$ is the adaptation parameter, balancing the two objectives that shape the features during training.  Upon training completion, the phase classifier is capable of predicting labels for both source and target domain samples.

\subsection{Model architecture and implementation}
Our optimized model employs four convolutional layers with hidden channels of 16, 32, 64, and 64, respectively. Each layer utilizes a 3$\times$3 convolutional kernel, a stride of 1, and padding of 2, followed by max pooling with the same kernel size and a stride of 2, and an activation function. All the activation functions used in this network are rectified linear units (ReLU) except for the first convolutional layer. To alleviate the dying ReLU problem where some neurons become permanently inactive and only output 0 for any input 
during training, the first convolutional layer ReLU is replaced with the leaky ReLU (LeakyReLU)\,\cite{maas2013rectifier} using a default constant slope of 0.01. Post-convolution, the network applies adaptive average pooling to the feature maps, which are then flattened to interface with the subsequent fully connected layers. The domain classifier and phase classifier are both two-layer fully connected neural networks with 64 hidden dimensions each. To enable phase label prediction, a softmax layer is added to the phase classifier to convert the predicted logits $\mu_{\alpha}$ for each phase $\alpha$ ($\alpha$ = SC or normal) into probability values $p_{\alpha}\propto e^{\mu_{\alpha}}$, obeying the sum rule $\sum_\alpha p_{\alpha}=1$. All models are implemented using PyTorch 1.12.1 and trained on a single NVIDIA RTX A5000 GPU. The optimized training procedure employs the Adam optimizer\,\cite{kingma2014adam} with a learning rate of 0.0005, weight decay (L2 penalty, 0.001), and early stopping. Dropout with a probability of 0.4 is applied after the last convolutional layer and in every layer of each fully connected neural network. Training is performed with mini-batches of 4 spectra for 150 epochs. The adaptation parameter is set to be 1.2 and is gradually increased from 0  through:
\begin{equation}
    \lambda=\frac{2}{1+e^{-\gamma\cdot p}} - 1\,,
\end{equation}
\noindent
in the early stages of the training, to suppress noisy signals from the domain classifier. Here, $\gamma$ is set to 10 without being optimized, and $p$ represents the training progress linearly changing from 0 to 1.

\subsection{Model evaluation and hyperparameter tuning}
Evaluating the performance of an unsupervised domain adaptation (UDA) model is challenging due to the lack of the target domain labels\,\cite{morerio2017minimal, saito2021tune, yang2023can}. As the domain classifier acts as a regularizer on CNN\,\cite{JMLR:v17:15-239}, we first optimize the hyperparameter of a CNN model without the domain classifier using simulated ARPES data (detailed in supplemental methods, Figure S3 and Table S5). We then fine-tune the adaptation parameter based on the transfer score (TS) metrics\,\cite{yang2023can}, which evaluates the transferability and discriminability of the feature space. The TS is formulated as follows: 
\begin{equation}
    \mathcal{T} = \mathcal{H} + \frac{|\mathcal{M}|}{\ln K}\,.
\end{equation}
\noindent
Here, $\mathcal{H}$ denotes the Hopkins statistic\,\cite{banerjee2004validating}, which measures the clustering tendency of the feature representation in the target domain; $\mathcal{M}$ denotes mutual information between the input and prediction in the target domain; $K$ is the number of classes for the normalization purpose. To obtain the Hopkins statistic, we define $\textbf{f}_t$ as the feature embeddings of all target domain samples, which is given by $\textbf{f}_t = [\mathrm{f}_1, \mathrm{f}_2, \ldots, \mathrm{f}_{n_t}]$, where $\mathrm{f}_j = G_f(\mathbf{x}_j^t; \theta_f)$.  From $\textbf{f}_t$, we randomly sample $m=0.05n_t$ data points\,\cite{lawson1990new}, without replacement to generate a set $R$. Additionally, we generate a set $U$ comprising $m$ data points sampled from a uniform distribution bounded by the minimum and maximum values along each feature dimension of $\textbf{f}_t$. We then compute two distance measures: $u_k$, the distance of samples in $U$ from their nearest neighbor in $R$, and $w_k$, the distance of samples in $R$ from their nearest neighbor in $R$. The Hopkins statistic is then defined as:
\begin{equation}
    \mathcal{H} = \frac{\sum^m_{k=1}u_k}{\sum^m_{k=1}u_k+\sum^m_{k=1}w_k}\,.
\end{equation}
\noindent
The mutual information is used to discern both the prediction confidence and diversity, as described in Yang et al.\,\cite{yang2023can}, and is defined as follows:
\begin{eqnarray}
    \mathcal{M} &=& H\left(\mathbb{E}_{\mathbf{x}_j^t \sim \mathcal{D}_t} G_y\left( G_f(\mathbf{x}_j^t;\theta_f);\theta_y \right) \right) \nonumber \\
    && - \mathbb{E}_{\mathbf{x}_j^t \sim \mathcal{D}_t} H\left(G_y\left( G_f(\mathbf{x}_j^t;\theta_f);\theta_y \right) \right)\,,
\end{eqnarray}
\noindent
where $H(\cdot)$ denotes the information entropy. With all the hyperparameter optimized, the model is then trained using all the labeled simulated training set and unlabeled experimental spectra. The model checkpoint with the best TS is collected for test set predictions.

\section*{Data Availability}
All the ARPES data that support the findings of this study are deposited in the Figshare repository ({\href{https://doi.org/10.6084/m9.figshare.25439632.v1}{https://doi.org/10.6084/m9.figshare.25439632.v1}}).

\section*{Code Availability}
The training script used to produce the findings of this study is publicly available via GitHub at {\href{https://github.com/Liu-group/ARPES}{https://github.com/Liu-group/ARPES}} and via Figshare at {\href{https://doi.org/10.6084/m9.figshare.25439623.v1}{https://doi.org/10.6084/m9.figshare.25439623.v1}}.

\section*{Acknowledgements}
We thank Mingda Li for insightful discussions. X.C. and F.L. acknowledge support by the U.S. Department of Energy, Office of Science, Basic Energy Sciences, under Early Career Award No.~DE-SC0025345. X.C., Y.S., and Y.W. acknowledge support from the Air Force Office of Scientific Research Young Investigator Program under grant FA9550-23-1-0153. Y.H. acknowledges support from NSF CAREER award No.~DMR-2239171. J.Y. was partially supported by a seed fund from the Yale Provost's Office. This research used resources of the National Energy Research Scientific Computing Center, a DOE Office of Science User Facility supported by the Office of Science of the U.S. Department of Energy under Contract No.~DE-AC02-05CH11231 using NERSC award BES-ERCAP0031226. 

\section*{Author contributions} 
Y.W., F.L., and Y.H. conceived the project. Y.H. and J.Y. prepared the dataset. X.C., Y.S., and E.H. built the ML model and carried out data analysis with the help of V.D. All the authors contributed to the interpretation of the results and the manuscript writing.

\section*{Competing interests} The authors declare no competing interests.

\bibliography{references}

\end{document}